\documentstyle[12pt,epsf,wrapfig]{article}
\newcommand{\degs}{^\circ}

\renewcommand\thesection{\arabic{section}.}
\renewcommand\thesubsection{\thesection\arabic{subsection}.}
\renewcommand\thesubsubsection{\thesubsection\arabic{subsubsection}.}
\renewcommand\section[1]{\vspace{\topsep}\vspace{\partopsep}
\refstepcounter{section}
{\par  \noindent\normalsize\bf \thesection
\hspace{1em}#1\vspace{\topsep}\par\noindent}}

\newenvironment{refs}
{\vspace{\topsep}\vspace{\partopsep}
{\par \noindent\normalsize\bf  References
\vspace{-\topsep}\par\noindent}
\setlength{\parindent}{-5mm}
\begin{list}{}{\topsep 0pt \partopsep 0pt \itemsep 0pt \leftmargin 5mm
\parsep 0pt \itemindent -5mm}}
{\end{list}}

\renewcommand\subsection[1]{
\refstepcounter{subsection}
{\par \protect\vspace{\topsep}\vspace{\partopsep}
 \noindent\normalsize\bf \it \thesubsection
\hspace{1em}#1\par \noindent}}

\renewcommand\subsubsection[1]{
\refstepcounter{subsubsection}
{\par \protect \vspace{\topsep}\vspace{\partopsep}
\noindent\normalsize \it \thesubsubsection
\hspace{1em}#1\par \noindent}}

\newfont{\sansb}{cmssbx10}
\newfont{\sans}{cmss10}

\begin{document}
\begin{center}
{\large \bf Stereo Imaging of VHE Gamma-Ray Sources\vspace{18pt}\\}
{F.A. Aharonian and A.K. Konopelko\footnote{Permanent 
address: Altai State University, Barnaul, Russia}\vspace{12pt}\\}
{\sl 
Max-Planck-Institut f\"{u}r Kernphysik, Heidelberg, Germany\\
}
\end{center}

\begin{abstract}

We describe the features of the stereoscopic approach in the
imaging  atmospheric Cherenkov technique, and discuss the 
performance  of future low threshold telescope arrays.

\end{abstract}


\setlength{\parindent}{1cm}
\section{Introduction}
The high detection rate, the ability of effective separation of 
electromagnetic and hadronic showers, and good accuracy of reconstruction
of air shower parameters are three remarkable features of the imaging atmospheric 
Cherenkov telescope (IACT) technique (see e.g. 
Cawley and  Weekes 1996,  
Aharonian and Akerlof 1997). The recent reports about detection of
TeV $\gamma$-rays from several objects by nine  groups
running imaging telescopes both in the northern (Whipple,
HEGRA, CAT, Telescope Array, CrAO, SHALON, TACTIC) and southern
(CANGAROO, Durham) hemispheres confirm the early expectations
concerning the potential of the technique (Weekes and  Turver 1977, Stepanian 1983, 
Hillas 1985), and provide a solid basis for the ground-based 
gamma ray astronomy (Weekes et al. 1997a). 

In particular, 
the Whipple 10~m diameter imaging telescope, today's most sensitive
single telescope,  provides detection of more than 100 
$\gamma$-rays from the Crab  with $\simeq 7$ standard deviation statistical
significance within only 1~h observation of the source. 
Thus the 100~h source exposure by this instrument
should be enough to reveal point $\gamma$-ray sources above 250 GeV
at the flux level of 0.07 Crab. On the other hand,  
the `7-sigma-per-1 hour' signal from the Crab  implies that 
any short outburst of a TeV source at the level exceeding
the Crab flux can be discovered during the observation time of less than 30 minutes.
This important feature of the IACT technique was convincingly demonstrated
by the Whipple group when two dramatic flares from Mrk 421
were detected in May 1996 (Gaidos et al. 1996). 
Significant improvement of the performance of
this telescope is expected in 1998,  after installation a  new,  
541-pixel high resolution camera (Lamb et al. 1995a).

While the cameras  with relatively modest pixel size
of about $0.25^{\circ}$ provide an adequate
quality of imaging of air showers 
produced by TeV $\gamma$-rays (Plyasheshnikov and Konopelko 1990, Zyskin et al. 1994, 
Aharonian et al. 1995),
a  smaller pixel size is preferable {\it (a)} for lowering 
the energy threshold, and {\it (b)} for extending  
the dynamical range towards the higher energies by observations of $\gamma$-ray sources 
at {\it large zenith angles}. Both  possibilities recently have 
been demonstrated respectively by CAT 
(Goret et al. 1997) and CANGAROO (Tanimori et al. 1997) groups.

Further qualitative  improvement of the IACT technique
in the next few years is likely to proceed in two (complementary)
directions: (a) implementation of the so called
{\it stereo imaging}, and (b) reduction of the
energy threshold towards the sub-100 GeV domain.

\begin{figure*}[htb]
\vspace{7.8 cm}
\includegraphics{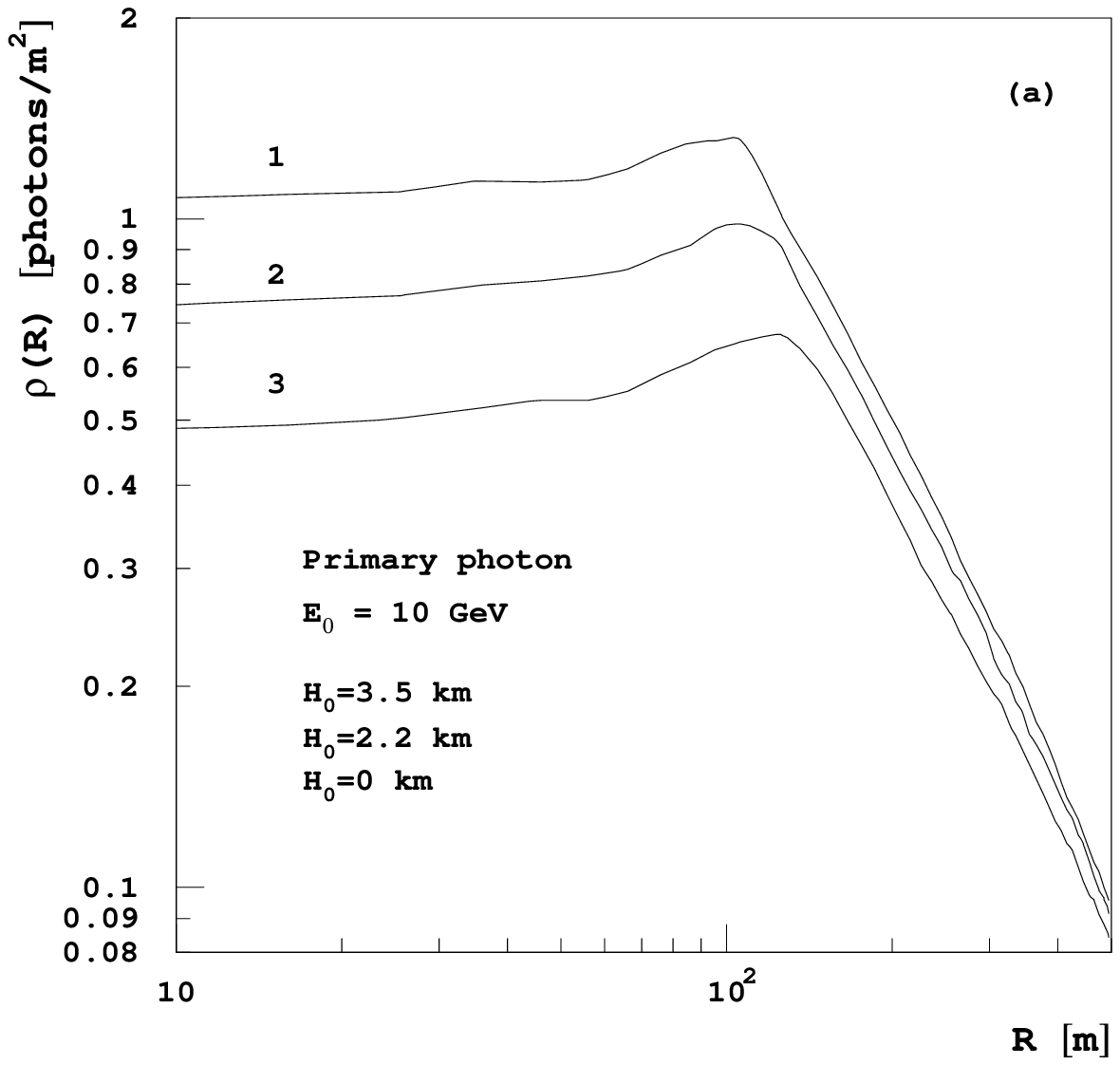}
\includegraphics{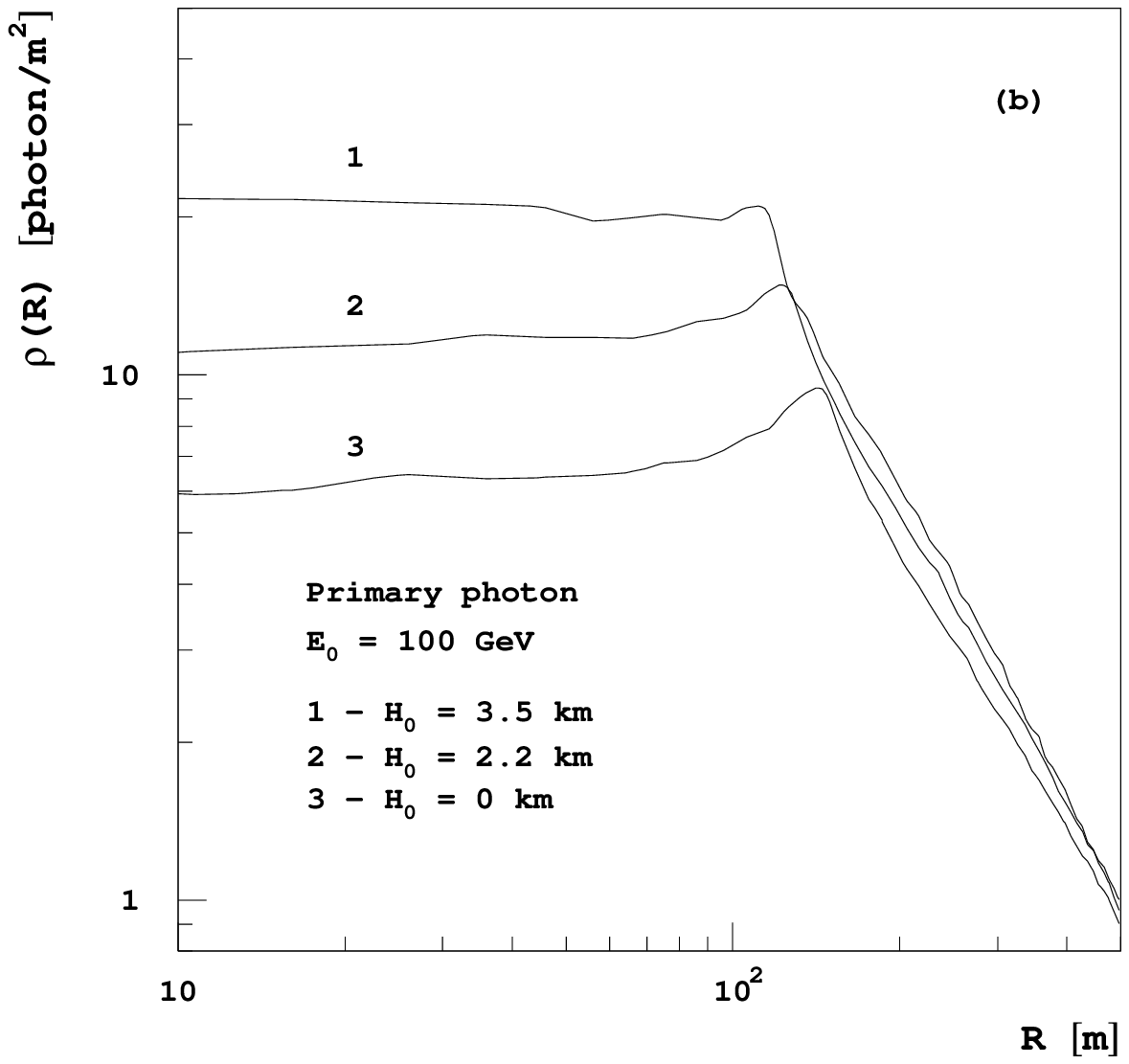} 
\caption{Density of Cherenkov light photons in 10~GeV and 100~GeV 
$\gamma$-ray showers as a function of the radial distance from 
the shower core.}
\end{figure*}
                                                               
\section{Stereoscopic Approach}

The concept of stereo imaging is based on the simultaneous detection
of air showers in different projections by at least two telescopes
separated at a distance comparable with the radius of the
Cherenkov light pool,  $R_{\rm C} \sim 100 \, \rm m$ (see Fig.~1).
Compared to single telescopes, the stereoscopic approach allows 
$(i)$ unambiguous and precise reconstruction of the shower parameters 
on an {\it event-by-event} basis, 
$(ii)$ superior rejection of hadronic showers, and 
$(iii)$ effective suppression of the background light of different 
origins -- night sky background ({\it N.S.B.}), local muons, {\it etc}
(Aharonian et al. 1993,  Aharonian 1993, Krennrich and Lamb 1995, Stepanian 1995,  
Chadwick et al. 1996, Aharonian et al. 1997).

The (only) {\it disadvantage} of the stereoscopic approach 
is a  significant loss of the detection rate due to the 
overlap of the shower collection areas of individual telescopes located 
from each other at distances $\leq 2 R_{\rm C}$. However, the loss
in  statistics is partially compensated,
especially for steep spectra of primary $\gamma$-rays,
by reduction  of the energy threshold of the telescopes 
operating in the coincidence mode.

While the effective collection area of a single telescope is determined 
by the radius of the Cherenkov light pool, the shower integration area 
of a multi-telescope system is due to the total geometrical area 
of the array which may be gathered from individually triggered groups of 
telescopes -- IACT {\it cells} (Aharonian et al. 1997). The compromise between 
two principal conditions, namely (a) detection of $\gamma$-ray induced showers 
by several telescopes of the {\it cell} and (b) minimization of correlation 
between images in different telescopes,  dictate the linear size of the 
{\it cell}: $L \simeq R_{\rm C}$. The design of the IACT {\it cell} depends
on the specific detection requirements. 
However, since the accuracy  of reconstructed shower parameters continues
to be improved noticeably up to 3 or 4 
telescopes in coincidence (Aharonian et al. 1997),
the optimum design of the {\it cell} seems to be a triangular or quadrangular
arrangement of IACTs. The concept of the IACT {\it cell} makes straightforward 
and rather general the predictions concerning   
the performance of multitelescope arrays.

\subsection{Energy Threshold}

\begin{wrapfigure}[19]{l}{9 cm}
\epsfxsize= 8 cm
\epsffile[0 15 300 300]{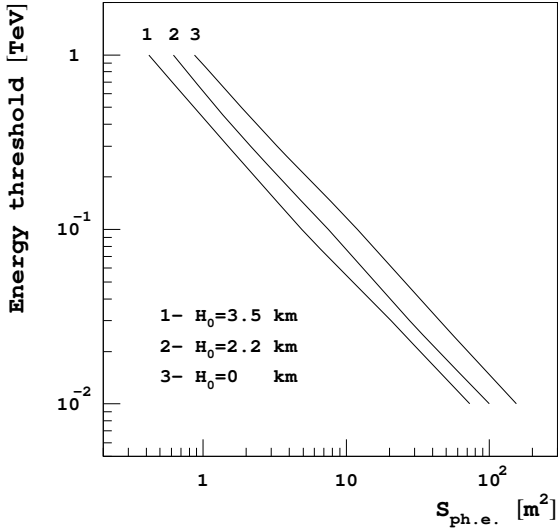}
\caption{Energy threshold of an imaging Cherenkov telescope versus 
the telescope's aperture $S_{\rm ph.e.}$.}
\end{wrapfigure}

The {\it effective energy threshold} of the imaging telescopes,
$E_{\rm th}$, is basicly determined by two conditions: $(i)$ the accidental 
trigger rate due to the {\it N.S.B.} should not exceed the detection rate of 
$\gamma$-rays, and $(ii)$ the number of photoelectrons in image should be sufficient 
for proper image analysis; typically $n^{\rm (min)}_{\rm ph.e.} \sim 100$.

Effective suppression of the {\it N.S.B.} in the IACT technique
is provided by the trigger criterion which requires signals above threshold 
in several adjacent pixels. Since the minimum number of pixels in coincidence
is limited by the characteristic image size, the high resolution cameras 
with pixel size of about $0.1^{\circ}$ have a certain advantage from the point 
of view of lowering the energy threshold  by using higher trigger multiplicity.

\begin{wrapfigure}[24]{l}{9 cm}
\vspace{7 cm}
\includegraphics{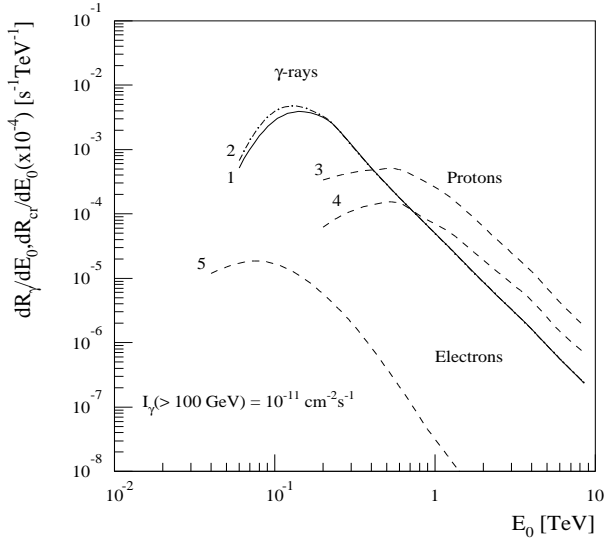}
\caption{Differential detection rates of $\gamma$-rays from a point source
with differential power-law index 2.5 and integral flux 
$J_\gamma(\geq 100 \, \rm GeV)=10^{-11} \, \rm ph/cm^2 s$, as well as
detection rates of isotropic cosmic ray protons and electrons ($\times 10^{-4}$) for the
4-fold telescope coincidence.  The curves 
2 and 4 are calculated for a camera with pixel size of $0.15^{\circ}$. All other 
curves are calculated for the pixel size $0.25^{\circ}$.
}
\end{wrapfigure}

\noindent
Meanwhile, the disadvantage of 
(more economical) cameras with relatively large (e.g. $0.25 \degs$) pixel size 
can be effectively compensated by the requirement of simultaneous detection of 
a shower by $\geq 2$ telescopes with a local  2-pixel coincidence trigger 
condition. 
Due to the flat lateral distribution of the Cherenkov radiation of 
$\gamma$-ray showers this requirement does not affect the 
$\gamma$-ray detection efficiency if the distance between telescopes
does not exceed $100 \, \rm m$.
Assuming now that the {\it local} and the {\it system} triggers eliminate 
effectively the  {\it non-air-shower} backgrounds, the threshold 
$E_{\rm th}$ could be estimated from the condition $(ii)$, 
namely from the equation 
$n_{\rm ph.e.}^{\rm (min)}=\rho(E) \times S_{\rm ph.e.}$,
where $\rho(E)$ is the density of the Cherenkov photons at plateau    
produced by a primary $\gamma$-ray of energy $E$, 
and $S_{\rm ph.e.}$ is the telescope aperture for the number of 
detected photoelectrons, 
$S_{\rm ph.e.} =S_{\rm mir} \cdot \chi_{\rm ph \rightarrow e}$. Here 
$S_{\rm mir}$ is the geometrical area of the optical reflector and 
$\chi_{\rm ph \rightarrow e}$ is the so-called `photon-to-photoelectron'
conversion factor, determined by quantum efficiency of the light detectors,
the reflectivity of the mirror, {\it etc}. If for order-of-magnitude estimates 
one assumes that $\rho(E) \simeq \rho_0 E$, the condition 
$n^{\rm (min)}_{\rm ph.e.}=100$ gives, for example at 2 km above see level, 
$E_{\rm th}=100 \, (S_{\rm ph.e.}/10 \, \rm m^2)^{-1} \, \rm GeV$.   
In Fig.~2 we show accurate calculations of the dependence of 
$E_{\rm th}$ on the telescope aperture $S_{\rm ph.e.}$ for different altitudes. 

From Fig.~2 one may conclude that the energy threshold  
around 100 GeV can be achieved, in principle, by a system of 10 m diameter 
IACTs equipped with conventional PMT based cameras with typical  
$\chi_{\rm ph \rightarrow e}=0.15$. The Monte Carlo calculations of detection 
rates for quadrangular 100~m $\times$ 100~m {\it cell} consisting of four IACTs
with aperture $S_{\rm ph.e.} = 10 \, \rm m^2$ show that 
the detection rate of $\gamma$-ray showers, with the cores inside the 
{\it cell}, indeed peaks at 100 GeV (Fig.~3). The efficiency of detection 
of 100 GeV photons even in the stringent 4-fold telescope 
coincidence mode is about $50 \%$, and reaches the $100\%$ level
already at 200~GeV.  

Note that $n^{\rm (min)}_{\rm ph.e.}=100$ is a rather
conservative condition. For relatively strong $\gamma$-ray sources, when 
a comprehensive image analysis is not required, the energy threshold
can be lowered down to 50 GeV.

\subsection{Reconstruction of Shower Parameters}

The procedure of reconstruction of the arrival direction and core position
of $\gamma$-ray showers by 4 telescopes of the {\it cell} is demonstrated 
in Fig.~4. Although one pair of detected images is sufficient to determine
unambiguously the shower parameters (except for the events close to the line 
connecting two telescopes), the information from other images significantly 
improves the accuracy of estimated shower parameters. 

\begin{wrapfigure}[19]{l}{8 cm}
\epsfxsize= 7.5 cm
\epsffile[5 100 560 650]{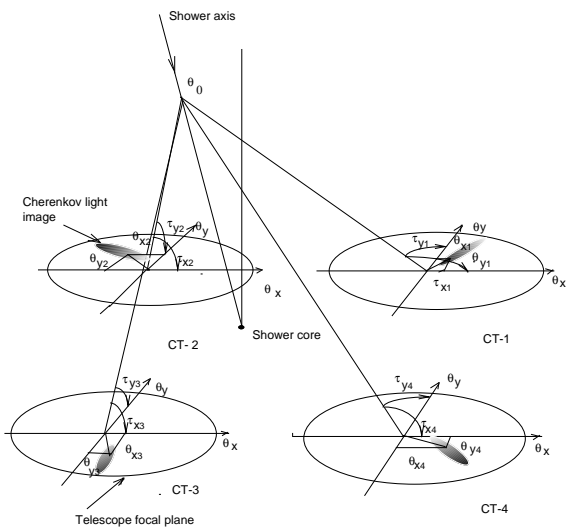}
\caption{
Reconstruction of the shower parameters by the 4-IACT {\it cell}.}
\end{wrapfigure}

The one dimensional angular distributions 
of showers produced by $\gamma$-rays from 
a point source with energy $E=$ 100~GeV, 300~GeV, and 1~TeV are shown in 
Fig.~5a. The two-dimensional distribution of the reconstructed  
directions of $\gamma$-rays from a point source with $E^{-2}$ spectrum is 
shown in Fig.~6. The values of the `$1 \sigma$' 
angular resolution  ($68 \%$ acceptance of $\gamma$-ray showers),  
are presented in Table~1. 

Detection of the shower images by $\geq 2$ telescopes allows precise 
determination of the shower core position (impact parameter). The 
distributions  of uncertainties in reconstructed impact parameter for 
different primary energies are  shown in Fig.~5b. The half width at half 
maximum (HWHM) values derived  from these distributions are presented 
in Table~1.

Determination of the shower core position with accuracy $\leq 10 \, \rm m$ 
is quite sufficient, due to the flat lateral distribution of the Cherenkov 
light from the  $\gamma$-ray showers (see Fig.~1), for the accurate evaluation 
of the energy using a relation between the primary energy $E$ 
and the total number of photoelectrons in the image. The accuracy of the 
energy estimate is essentially due to  the fluctuations which are large 
($\geq 30 \%$) for very close ($R \leq 60 \, \rm m$) as well as distant 
($R \geq 120 \, \rm m$) showers (see Konopelko 1997). However, most 
of the showers required to be detected by all four telescopes are located 
between 60~m and 120~m from at least 2 telescopes of the {\it cell}. 
Therefore the fluctuations appear to be small enough to provide determination
of the energy of $\gamma$-ray showers with an accuracy of about 
$20 \%$ in the broad energy range from 100~GeV to 1~TeV (see Table~1).   

\begin{figure*}[t]
\vspace{7.3 cm}
\includegraphics{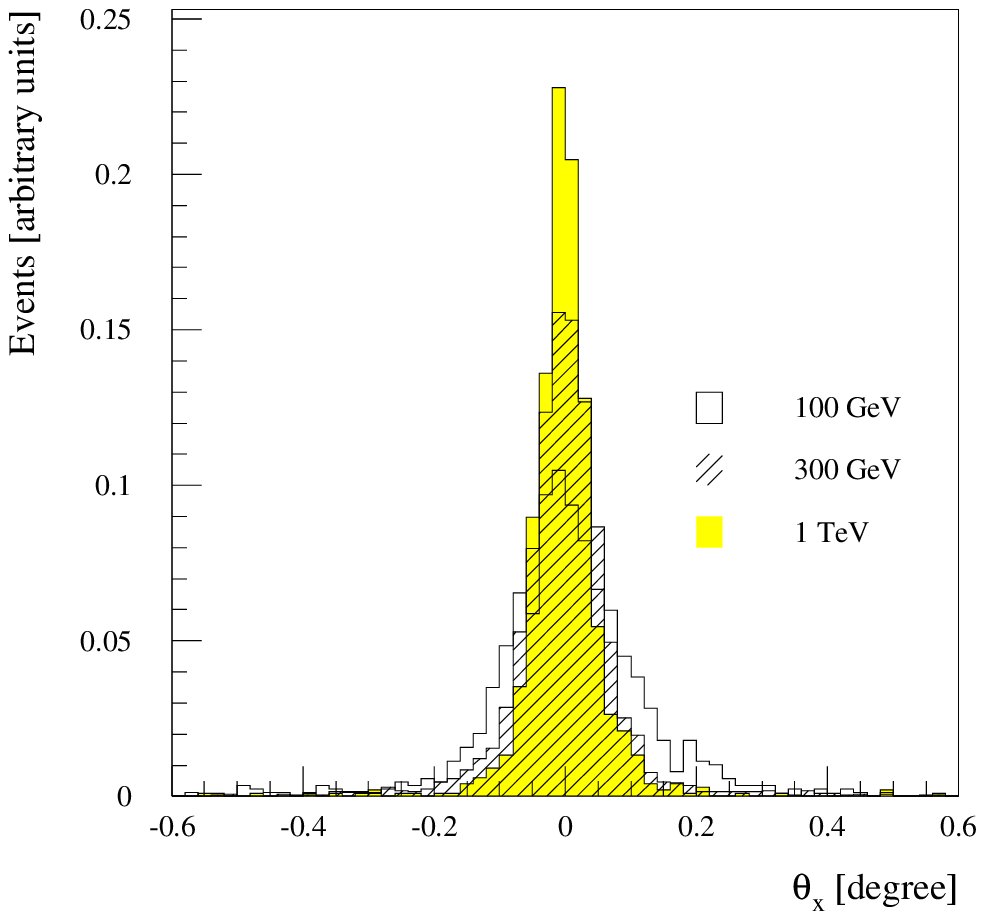}
\includegraphics{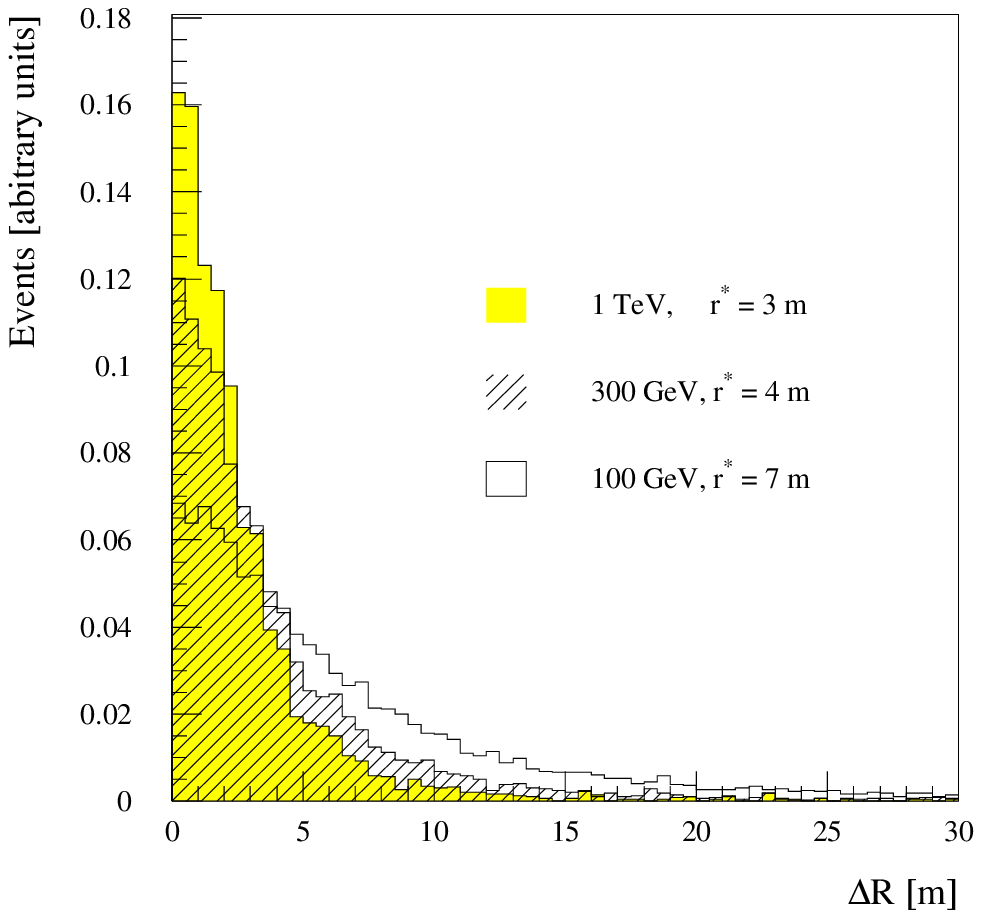}
\caption{a (left):~One-dimensional angular distribution 
of $\gamma$-ray showers.~ b (right):~Uncertainties in the reconstructed impact 
parameter of $\gamma$-ray showers.}
\end{figure*}

\subsection{Gamma/Hadron Separation}

\begin{wrapfigure}[20]{l}{8.5 cm}
\epsfxsize= 6.8 cm
\epsffile[0 20 290 290]{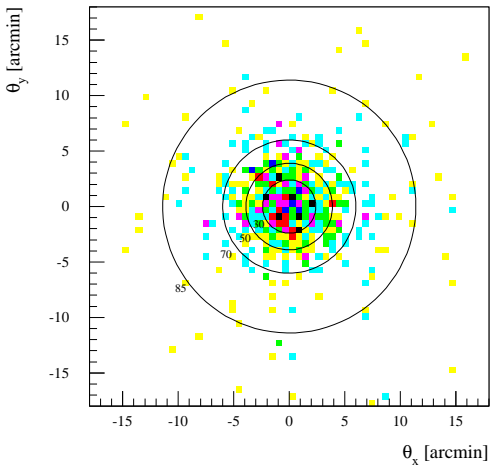}
\caption{Two dimensional distribution of reconstructed arrival directions
of $\gamma$-rays from a point source with the differential spectrum 
$dJ_\gamma/dE \propto E^{-2}$.
Circles correspond to the $30 \%, \, 50 \%, \, 70 \%, \, 85 \% \, $ $\gamma$-rays acceptance.}
\end{wrapfigure}

Determination of the arrival direction of individual $\gamma$-ray showers with 
an accuracy $\leq 0.1^{\circ}$ allows effective suppression of the cosmic ray 
background for  point sources by a factor of 
$\kappa_{\rm cr}^{\rm (dir)}=(2 \delta \theta/Psi)^2$,   
where $\Psi$ is the  the angular diameter of the `trigger zone'. 
For point sources typically $\Psi \sim  3^{\circ}$, thus  
$\kappa_{\rm cr}^{\rm (dir)} \simeq 1/300$.
Further improvement of the signal-to-noise ratio is provided by
exploitation of the intrinsic differences between the development of 
the electromagnetic and hadronic cascades. These  differences, caused by 
the large transverse momenta of the secondary hadrons, the deeper penetration 
of the hadronic cascades into the atmosphere, and essential fluctuations 
in the hadronic cascades (e.g. Hillas 1996), are effectively transferred to 
to the shape of the 
Cherenkov light images. The  $\gamma$-ray images  
have compact shape and regular structure, in contrast to rather 
diffuse and chaotic images of the hadronic showers. For {\it single} 
telescopes this provides rejection  of hadronic showers,
based on the
so-called {\it shape} cuts, by a factor of 10 or so (see e.g. Fegan 1997).

\begin{figure*}[t]
\vspace{7.3 cm}
\includegraphics{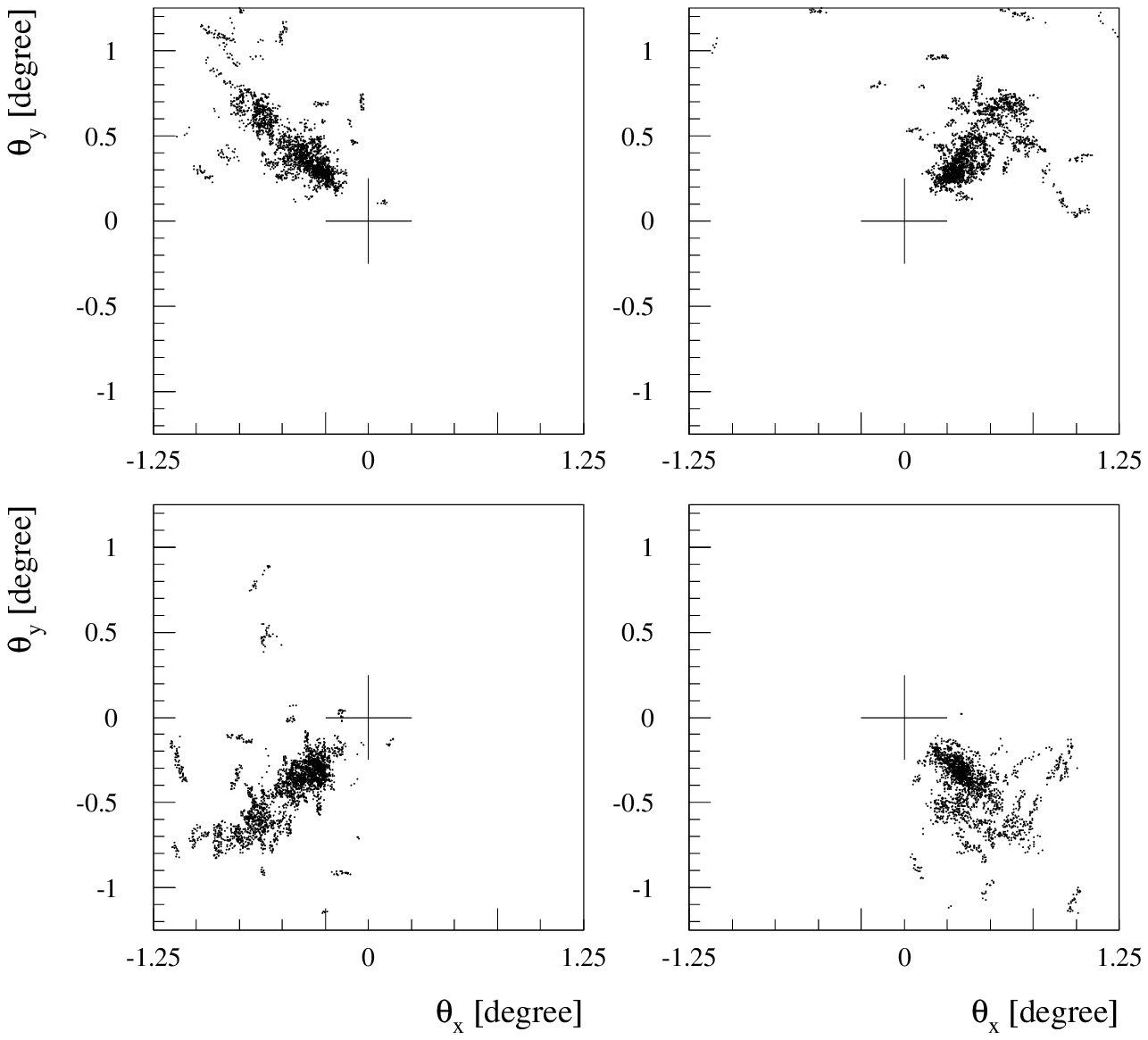}
\includegraphics{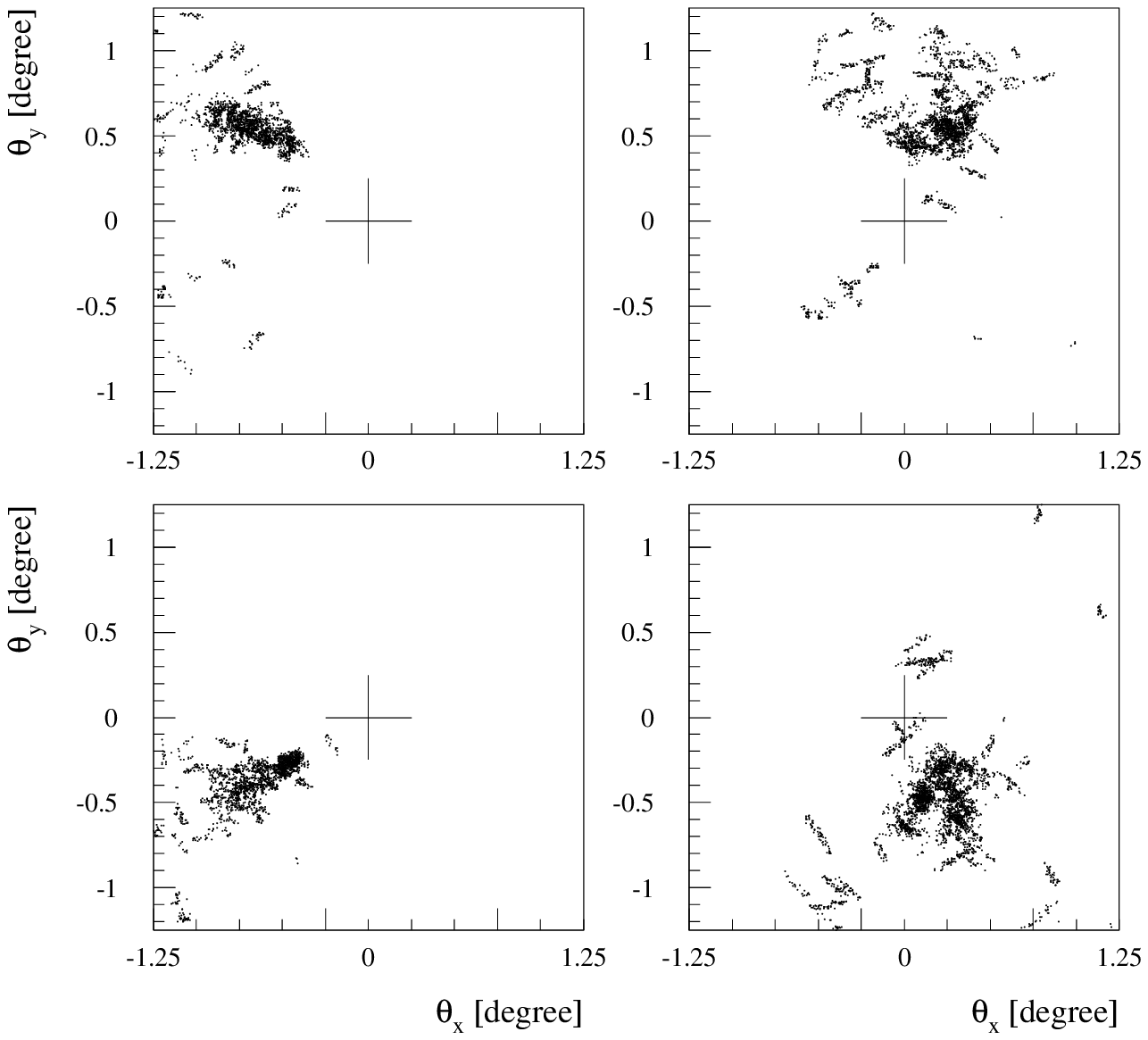}
\caption{
Cherenkov light images of vertical 100-GeV $\gamma$-ray
(left) and 300-GeV proton (right) detected by four telescopes of a 
100 m $\times$ 100 m IACT cell. Both showers cross the center of the cell. 
While the images of the proton shower in two telescopes look like 
$\gamma$-ray showers (narrow and regular), the images obtained by two 
other telescopes clearly indicate the hadronic nature of the shower.}
\end{figure*}

Determination of the arrival direction, energy and core location of showers 
on an {\it event-by-event} basis with accuracy $\leq 20 \%$, 
$\leq 0.1^{\circ}$, and $\leq 10 \, \rm m$, respectively, allows
an optimization of the image `cuts'  which separate the CR and $\gamma$-ray
domains in the shower parameter space, and thus improves
the rejection efficiency for hadronic showers. 
Furthermore, even though the images in different projections are not
entirely independent of each other, the correlation is only partial.
Therefore, the joint analysis dramatically improves the
gamma/hadron separation power (this effect is demonstrated in Fig.~7). 
The characteristic values of the acceptance of $\gamma$-ray ($\kappa_\gamma$)
and proton ($\kappa_{\rm h}$) induced showers, as well as the so-called
Q-factor, $Q=\kappa_\gamma/\kappa_{\rm h}^{1/2}$ (the enhancement
of the signal-to-noise ratio after the application of the cuts;
see, e.g. Fegan 1997), are presented in Table 2.

%
\begin{table}[h]
\caption{The angular, distance, and energy resolutions of the 4-IACT {\it cell}.}
\begin{center}
\begin{tabular}{|l|l|l|l|} \hline
                        & 100 GeV & 300 GeV & 1 TeV \\ \hline
$\sigma_\theta$, arcmin & 6       & 4.5     & 3     \\ \hline
$\delta r$, m           & 3       & 4       & 7     \\ \hline
$\delta E/E$, \%          & 20      & 20      & 22    \\ \hline
\end{tabular}
\end{center}
\end{table}

It is seen that rejection efficiency of hadronic showers significantly 
increases with the number of detected images, and under the condition of 
$50 \%$ acceptance of $\gamma$-ray events the rejection factor $\kappa_{\rm h}^{-1}$ 
of 4-IACT 
{\it cell} can be as large as 300. This results in the Q-factor of 
about 8. Note that such effective gamma/hadron separation is provided by 
application of a `scaled' {\it Width} parameter (Konopelko~1995). 
The  other parameters, like image {\it Length}, time structure, UV-content 
of the Cherenkov light flashes, {\it etc.}, are less effective, but they still 
can improve the efficiency of suppression of the hadronic showers being 
applied separately. However, in the case of stereoscopic observations, 
the {\it Width} cuts in different projections are so effective, that they  
practically do not leave a room for further enhancement of the signal 
(Aharonian et al. 1997).  

\begin{figure*}[t]
\vspace{5.8 cm}
\includegraphics{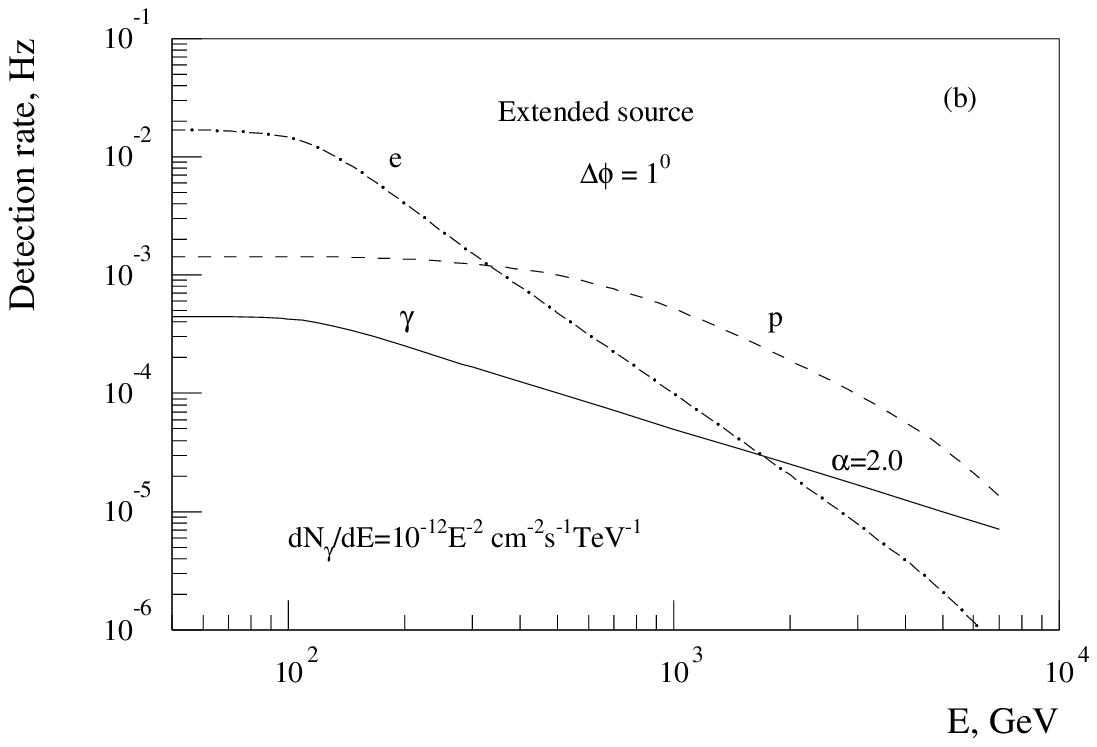}
\includegraphics{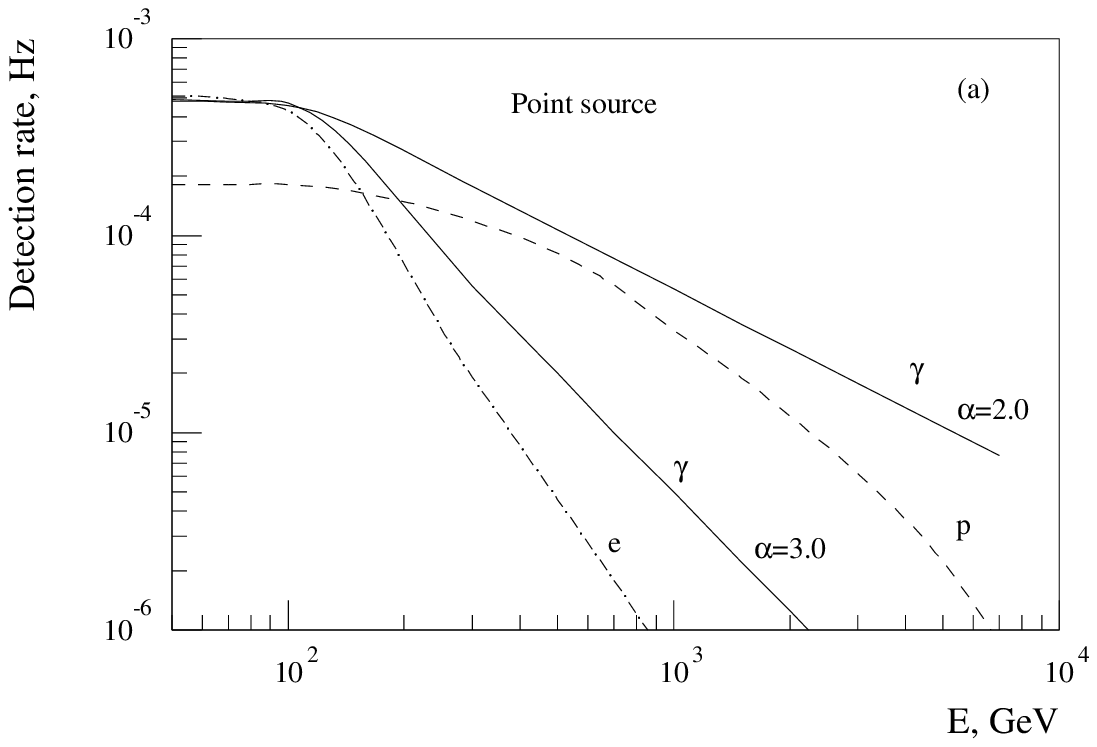}
\caption{Detection rates of $\gamma$-rays (solid lines), as well as cosmic ray 
protons (dashed) and electrons (dot-dashed)
by the  $100$~m $\times$ $100$~m IACT {\it cell}, after application 
of software image cuts.
(a) -- point source, and  (b) -- `$1 \degs$' diffuse source of $\gamma$-rays.
In both cases an integral flux of 
$\gamma$-rays $J_\gamma(\geq 100 \, \rm GeV)=10^{-11} \, \rm ph/cm^2 s$ with 
power-law indices $\alpha=2$ and $\alpha=3$ is assumed. 
In the case of point sources the combined $50 \%$ acceptance for $\gamma$-rays
is achieved by applying directional and {\it relaxed}  shape cuts, while in 
the case of a `$1^{\circ}$' source the same $50 \%$ $\gamma$-ray acceptance
is a result of {\it tight} shape cut.}   
\end{figure*}

%
\begin{table}[b]
\caption{CR background rejection efficiency of IACT `cell'
using parameter {\it Width} with and without
`scaling'. In the analysis the N-fold {\it Width} cuts were used 
(N = 2,3,4; N is the number of the images passed the cut).}
\begin{center}
\begin{tabular}{|l|l|l|l|l|l|} \hline
Parameter:  & Scaling: &  & 2 & 3 & 4 \\ \hline
            &          & $\kappa_\gamma$ & 0.792 & 0.585 & 0.489  \\ 
{\bf W}idth & No       & $\kappa_{\rm h}$   & 0.116 & 0.038 & 0.016  \\ 
            &          & Q-factor        & 2.32  & 3.00  & 3.87   \\ \hline
            &          & $\kappa_\gamma$ & 0.891 & 0.634 & 0.490  \\ 
{\bf W}idth & Yes      & $\kappa_{\rm h}$   & 0.076 & 0.017 & 0.0036 \\ 
            &          & Q-factor        & 3.23  & 4.86  & 8.33   \\ \hline
\end{tabular}
\end{center}
\end{table}

The suppression of the cosmic ray background based on the {\it directional} 
and {\it shape} criteria for point sources could be as effective 
as $\kappa_{\rm cr}=\kappa_{\rm cr}^{\rm (dir)} \kappa_{\rm h}
\simeq  10^{-5}$ with the Q-factor close to 100. 
For comparison, CR rejection efficiency and Q-factor in case of single 
IACTs, after the application of both {\it orientational} ({\it Alpha})
and {\it shape} cuts, hardly can exceed $10^{-3}$ and 10, respectively.     

Due to the high rejection efficiency of hadronic showers by the IACT {\it cell}, 
the cosmic ray background at energies below 200 GeV is dominated by showers 
produced by electrons (see Fig.8~a,b). This allows   
detection of $\gamma$-rays from point sources under almost 
{\it background-free} conditions if the photon flux above 100~GeV exceeds 
$10^{-11} \, \rm ph/cm^2 s$ (approximately 1/20 of the Crab flux expected at
these energies).

\section{HEGRA: The Stereo Imaging Does Work !}

\begin{wrapfigure}[20]{l}{8.5 cm}
\epsfxsize= 7.5 cm
\epsffile[73 525 356 793]{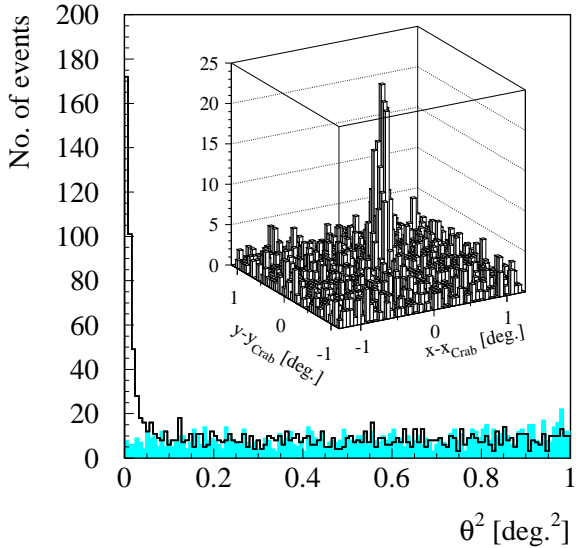}
\caption{Distribution of the reconstructed directions of the showers detected 
from the Crab by the  HEGRA IACT system  (from Daum et al. 1997).}
\end{wrapfigure}

The recent observations of the Crab Nebula and Mrk~501 by HEGRA stereoscopic 
system of 4 IACTs (Daum 1997, Herman 1997, Hofmann 1997)  
confirm the early predictions for the performance of the instrument 
(Aharonian 1993). Fig.~9 shows the angular
distribution of showers detected from the direction of the Crab,
and selected after the image {\it shape} cuts (Daum et al. 1997).
The confinement of the $\gamma$-ray signal
from a point source in a very small angular region (with radius
$\sim 0.1^{\circ}$) of the available two-dimensional phase space,
in addition to the significant suppression of hadronic showers at the 
{\it trigger} level, results in a strong $4 \sigma$-per-1~h signal already 
before the gamma/hadron separation. The {\it shape} cuts provide further, by a 
factor of 100, suppression of the hadronic background, while maintaining the 
efficiency of the acceptance of $\gamma$-ray events at the level of $50 \%$. 
This leads to only $\simeq 1$ background event in the `signal' region while 
the rate of $\gamma$-rays from the Crab exceeds 20 events/h. This enables 
$(i)$ an effective search for VHE $\gamma$-ray point sources with fluxes down 
to 0.1~Crab at almost {\it background free} conditions, and thus allows drastic
reduction of the observation time (by a factor of 10) compared to four 
independently operating telescopes, $(ii)$ detailed spectroscopy of strong 
(Crab-like) VHE emitters, $(iii)$ study of the spatial distribution of 
$\gamma$-ray production regions on arcminute scales, $(iv)$ effective search 
for {\it extended} sources with angular size up to $\simeq 1^{\circ}$ at the flux 
level of 0.1~Crab. The exploitation of the `nominal' 5-IACT HEGRA system with 
improved trigger condition should allow $5 \sigma$ detection of faint 
$\gamma$-ray sources at the flux level of 0.025~Crab 
(Aharonian 1997, Hofmann 1997) 
which corresponds to $\simeq 10^{-12} \, \rm erg/cm^2 s$ energy flux above 
the effective energy threshold of the instrument of about 500 GeV.

\section{Future 100 GeV Class IACT Arrays}

One of the principal issues for future detectors is the choice of the energy 
domain. If one limits the energy region
to a relatively modest threshold around 100 GeV, the performance
of IACT arrays and their practical implementation
can be prediced with high confidence. In practice, an energy threshold of 
100 GeV can be achieved by a stereoscopic system of IACTs 
consisting of 10~m diameter optical reflectors   
equipped with conventional PMT-based high resolution cameras.
Two such arrays of 10~m class telescopes -- 
VERITAS (Weekes et al. 1997b) and HESS (Hofmann 1997) -- are likely to be constructed 
in the foreseeable future.

\begin{figure*}[htb]
\vspace{15.5 cm}
\includegraphics{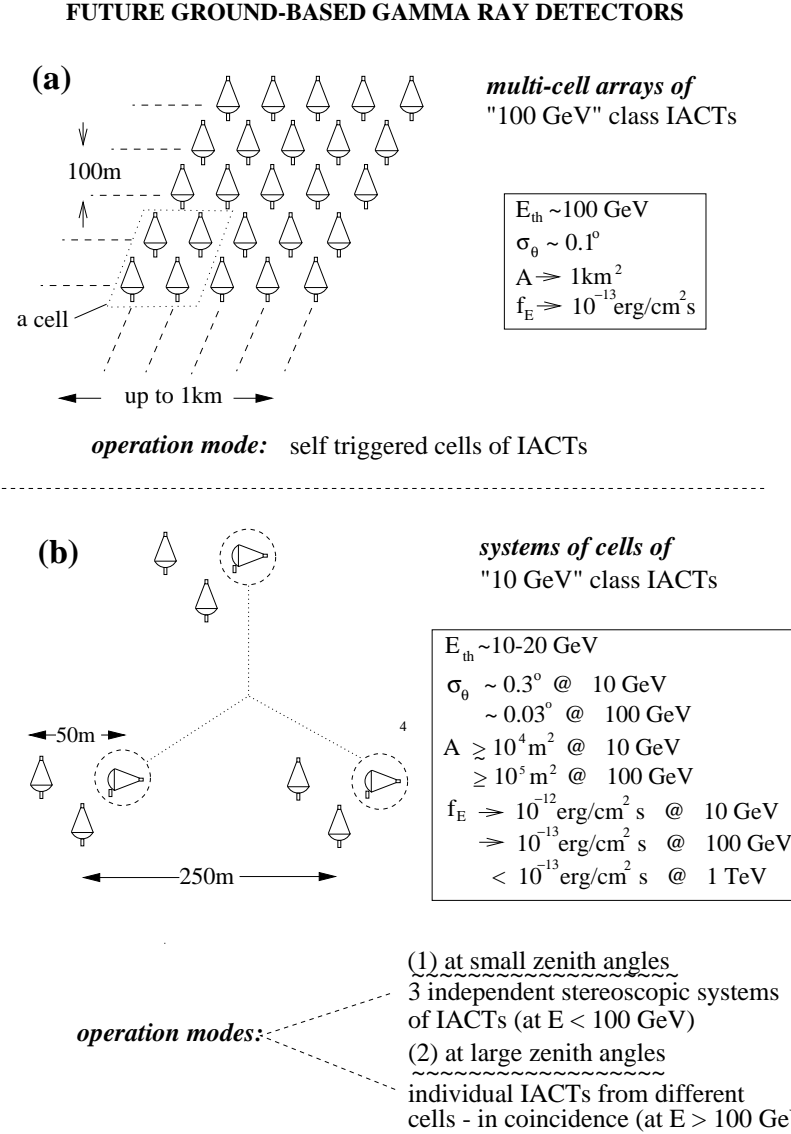}
\caption{Two possible arrangements of (a) 100~GeV class and 
(b) 10~GeV class IACT arrays.}
\end{figure*}

After reaching the maximum possible suppression of the cosmic ray background by
simultaneous detection of the air showers in different projections -- limited
basically by intrinsic fluctuations in cascade development -- the further
improvement of the flux sensitivities for a given energy threshold
can be achieved by increasing  the shower collection area,
i.e. the number of IACT {\it cells}.
A possible arrangement of a cell-structured array is schematically shown in Fig.~10a. 
For the arrays consisting of a large number of {\it cells}, e.g. $n_0 > 10$,
and for a given observation time $t_0$,
the minimum detectable flux of $\gamma$-rays from a source of angular size $\phi$
is determined essentially by the product $A=n_0 \times t_0$ from the following
two conditions: (1) high confidence level of a $\gamma$-ray signal, i.e. 
$\sigma=R_\gamma (R_{\rm p}+R_{\rm e})^{-1/2} A^{1/2} G^{-1}(\phi)  \geq \sigma_0$,
and (2) adequate photon statistics, i.e. $N_\gamma=R_\gamma A \geq N_0$.
Here $R_\gamma$, $R_{\rm p}$, and  $R_{\rm e}$ are the detection rates of 
$\gamma$-rays from a point source, and the detection rates of cosmic ray protons 
and electrons by one IACT {\it cell}, respectively  (see Fig.~8a). 
The parameter $G(\phi) \simeq \rm max[1, \phi/\delta \theta]$ 
takes into account the angular size of the source.
In Fig.~11 we show the minimum detectable $\gamma$-ray fluxes from a 
point source and $1^{\circ}$
source requiring $\geq 5 \sigma$ appearance of a signal,  provided that the
number of detected $\gamma$-rays exceeds 25. Note that at a large 
exposure, e.g. $A \geq 2500 \, \rm h$, which can be realized, for example,
by 100~h observations of a source with an array consisting of 
25 IACT-{\it cells}, 
$J_\gamma^{\rm min} (\geq 100 \, \rm GeV) \sim 10^{-12} \, \rm ph/cm^2 s$, or 
the corresponding 
energy flux $f_{\rm E} \sim 10^{-13} \, \rm erg/cm^2 s$. 

\begin{figure*}[t]
\vspace{9 cm}
\includegraphics{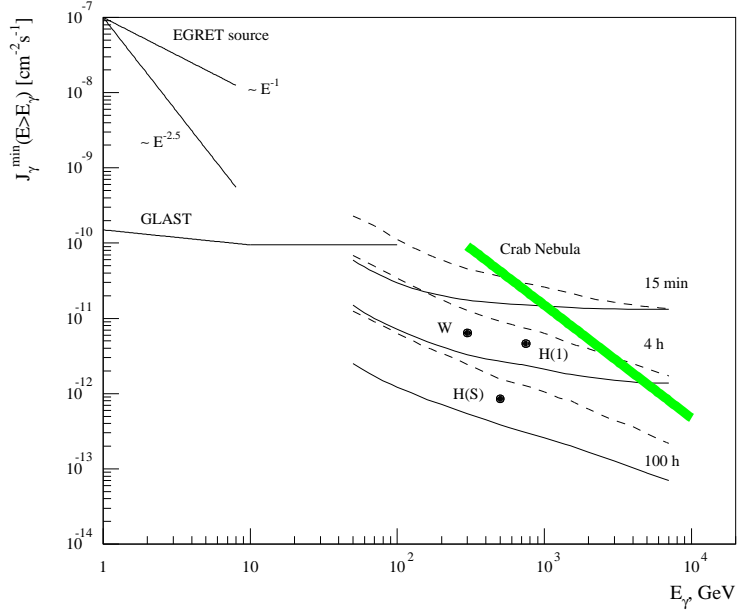}
\caption{
Minimum detectable fluxes of $\gamma$-rays from a point 
source (solid lines) and $1^\circ$ source (dashed lines) within 100~h 
observation time by an array consisting
of twenty five 10~m diameter IACT {\it cells}. The calculations correspond to the 
requirement of a 5$\sigma$ $\gamma$-ray signal with at least 25 detected 
photons. For comparison,  we show the following fluxes:
$(i)$ power-law extrapolations of a `standard'  EGRET source 
with integral 
indices $(\alpha - 1) =$ 1 and 2.5; $(i)$ 
the level of the measured $\gamma$-ray fluxes of the Crab Nebula  
above 300 GeV, assumed in the form of 
$\rm J_{\rm Crab}(\geq E) \simeq 1.5 \times 10^{-11} \, (E/1 \, TeV)^{-1.5} \, ph/cm^2 s$. 
The minimum detectable fluxes of $\gamma$-rays by the single HEGRA and 
Whipple telescopes (indicated as H(1) and W, respectively), and    
by the HEGRA IACT system (H(S)), calculated for 100~h 
observations of a point source,  are also shown.}
\end{figure*}

At a small exposure, e.g. $A=n_0 \times t_0  \leq 100 \, \rm h$, the minimum detectable 
fluxes are limited, especially at high energies,  
by low $\gamma$-ray statistics. This could be treated as a 
disadvantage of cell-structured arrays with fixed linear size of the {\it cell}, 
since  such arrays with $L \sim R_{\rm C}  \sim 100 \, \rm m$
are optimized, to a large extent, to the energy region close to  the threshold.
Indeed, although  the homogeneous multi- cell arrays
provides rather economical coverage
of large detection areas 
at the threshold energies
(each telescope is exploited by  four {\it cells}),
at high energies this  concept becomes less effective. 
Namely, it cannot use the advantage that the showers initiated by high energy photons,
$E \gg E_{\rm th}$, can be detected at distances well beyond 100~m. 
However this disadvantage of homogeneous cell-structured arrays can be 
compensated, at least partially, by observations at large zenith angles 
by $\geq 2$ telescopes in coincidence from different peripheries of the array.

\begin{wrapfigure}[19]{l}{9 cm}
\epsfxsize= 8 cm
\epsffile[0 20 350 300]{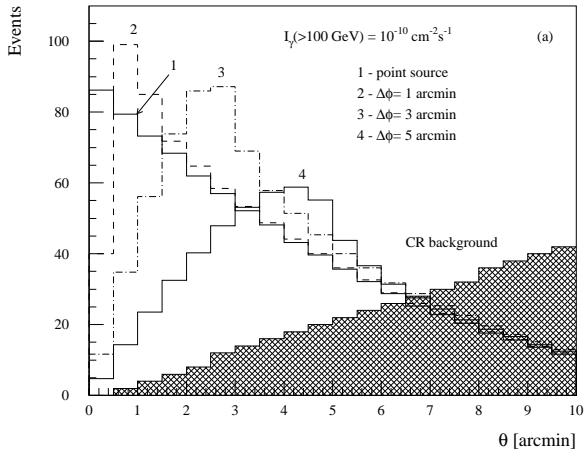}
\caption{Distributions of the reconstructed arrival directions of 
$\gamma$-rays from sources with different angular radius.
The histogram 1 corresponds to the two-dimensional
distribution shown in Fig.~6.} 
\end{wrapfigure}

From this point of view it would be important to equip some of the telescopes of the array,
located at large distances from each other, by very high resolution cameras with pixel
size $\simeq 0.1^{\circ}$.   
 
The determination of arrival directions of $\gamma$-rays by IACT arrays 
on an {\it event-by-event} basis with accuracy
$\leq 0.1^{\circ}$ makes possible a study 
of the spatial distribution of VHE $\gamma$-ray sources
in arcminute scales. This is demonstrated in Fig.~12, where the distributions of 
the angular distance of reconstructed directions of $\gamma$-rays 
from the center of extended source are shown. 
From this figure we may conclude that for sufficiently high photon statistics 
(say $\geq 10^3$ detected $\gamma$-rays) it would be possible to 
measure the size
of an extended source of $\phi \leq 0.1^{\circ}$ with  accuracy $\leq 1$ arcmin.

Good energy resolution of about 
$\delta E/E \sim 20 \%$  enables reasonable spectrometry with 
IACT arrays with ability, for example, to reveal spectral features like sharp 
cutoffs in the  spectra of VHE sources (see Fig.~13).

Due  to very effective rejection of the hadronic showers 
by the stereoscopic IACT systems, the cosmic ray background  
at energies close to 100 GeV is dominated by the showers produced by primary electrons.
This component of electromagnetic showers 
remains as a part of the background which principally cannot be removed, and thus 
it is the most serious limiting factor of flux sensitivities, especially for extended sources.
On the other hand, these electromagnetic showers with known flux and spectrum of
the primary electrons, measured 
up to energies 1 TeV (Nishimura et al. 1980), can be used for 
absolute {\it energy calibration} of the IACT arrays. Although the current uncertainties
in 100~GeV electron fluxes could  be as large as factor of 2, 
the future balloon and satellite measurements (e.g. Torii et al. 1997, 
M\"uller et al. 1997,  Ormes et al. 1997),
should be able to reduce significantly the flux uncertainties. 

The energy calibration by using the cosmic ray electrons provides unique tool for
the {\it continuous} (on-line) control of the characteristics of 
IACT arrays (e.g. the energy threshold, 
the detection area, {\it etc.}) during the $\gamma$-ray  observations. 
For example, 10~min exposure will be enough for 
detection of more than 100 electrons within the field of view of 
100~GeV threshold array consisting of 9 IACT {\it cells} 
(see Fig.~8b). 
It is difficult to overestimate the significance of such 
calibration and control, especially for the study of
the spectral characteristics of highly  
variable $\gamma$-ray sources on sub-hour timescales.   

\section{Sub-100 GeV Ground-Based Detectors}

The strong scientific motivations to fill the gap between
the space-based and ground based observations
recently stimulated several interesting proposals for extending the
atmospheric Cherenkov technique down to 10~GeV
region (see e.g. Lamb et al. 1995b). Among a variety of
competing projects  this goal
could be addressed by the
stereoscopic systems of IACTs with  the telescope apertures
of $S_{\rm ph.e.} \geq 50 \, \rm m^2$.
With development of the technology of novel, fast optical
detectors of high quantum efficiency
($\chi_{\rm ph \rightarrow e} \geq 50 \%$) and
design of (relatively inexpensive) 20 m class reflectors with
an adequate optical quality for the Cherenkov imaging on scales
of several arcminutes,  it would be possible to
reduce the detection threshold down to 20 GeV,  or even 10 GeV
for systems installed at high mountain
elevations (e.g. 3.5 km a.s.l).
Considerable R\&D  efforts, in particular by the
Munich group (Lorenz 1997, Mirzoyan 1997) are already started
in both directions. It is expected that this
activity  will result in construction of a large single reflector
telescope (MAGIC)  which, together with two ongoing
projects of low  energy threshold air Cherenkov detectors based on
large mirror assemblies of the existing solar power plants --
STACEE (Ong et al. 1997) and CELESTE (Par\`{e} et al. 1997) --
will start to explore the energy region below  100 GeV.
Moreover, if successful, the MAGIC telescope could be considered
as a prototype for the basic element in
future (post `VERITAS/HESS')  arrays  of 10 GeV class IACTs.

\begin{wrapfigure}[30]{l}{9 cm}
\epsfxsize= 8.3 cm
\epsffile[0 20 300 325]{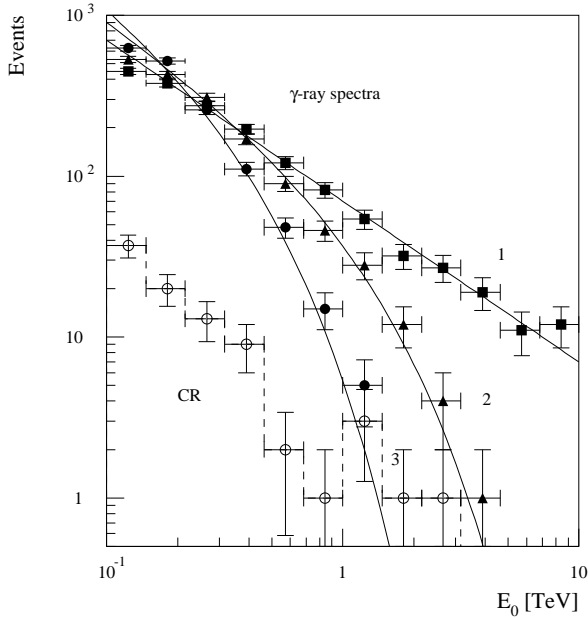}
\caption{Evaluation of differential spectra of $\gamma$-rays detected by an array 
for $A=n_0 \times t_0=50 \, \rm h$. 
The primary spectra (solid lines) have a power-law form $\propto E^{-2.2}$ with 
different values of a cutoff energy:
1 - $E^\star > 10 \, \rm TeV$ (corresponding reconstructed fluxes are shown by 
filled quadrangles);
2 - $E^\star  = 1 \, \rm TeV$ (reconstructed fluxes - filled triangles);
3 - $E^\star = 300 \, \rm GeV$ (reconstructed fluxes - filled circles).
The CR background events, remaining after application of
image `cuts', are shown by open circles.}
\end{wrapfigure}

The efficiency of the imaging technique
at energies well below 100~GeV seems to be
lower than at higher energies. Indeed, at energies close to  
10~GeV, the $\gamma$-ray images become less
elongated (more circular) and less regular.
In practice this means an introduction of
significant uncertainties in the reconstruction
of the arrival direction of $\gamma$-rays, as well as in the
gamma/hadron separation strategies. In particular,
our preliminary results show that the angular resolution of
stereoscopic systems at 10 GeV hardly can be better than $0.2^{\circ}$.
Also, due to large fluctuations in low-energy hadronic showers
(especially due to the higher probability of
production of a leading $\pi^{0}$ meson which takes almost all energy 
of the primary hadron), an essential fraction of
the cosmic rays could ``pass'' the {\it shape} cuts, i.e.
could be accepted  as $\gamma$-rays.
On the other hand, since the electron/proton ratio of cosmic rays
increases at lower energies as $\propto E^{-0.6}$, the weakened rejection efficiency
of hadronic showers seems to be still sufficient for operation
of the  {\it stereoscopic} sub-100 GeV IACT systems in the regime
when the background is dominated
by the cosmic ray electrons. This makes straightforward the prediction
for the flux sensitivity of such systems. Indeed, since
the showers caused by electrons are quite similar to the $\gamma$-ray showers,
the condition of detection of $5 \sigma$ signal of $\gamma$-rays with
differential spectrum ${\rm d} J_\gamma/{\rm d} E$ at any energy $E$ within
the interval $[E - \Delta E, E + \Delta E]$ results in
$$\left(\frac{{\rm d}J_\gamma}{{\rm d} E}\right)_{\rm min}=5 \,
\left (
\frac
{({\rm d} F_{\rm e}/{\rm d}E {\rm d} \Omega) \Delta \Omega}
{2 \Delta E t_0 S_{\rm eff} (E) \kappa_{\rm e.-m.} \kappa_\theta^2}
\right )^{1/2},
$$
where ${\rm d} F_{\rm e}/{\rm d}E {\rm d} \Omega \simeq
4 \cdot 10^{-2} E^{-3.2} \, \rm cm^{-2} s^{-1} sr^{-1} GeV^{-1}$ is the
measured spectrum of the cosmic ray electrons;
$S_{\rm eff}(E)$ is the effective area of
detection of electromagnetic showers,  
$\kappa_\theta$ and $\kappa_{\rm e.-m.}$ are the acceptance of electromagnetic
showers after application of the
{\it directional} and {\it shape} cuts;
$\Delta \Omega \simeq \pi \sigma_{\theta}^2$
is the solid angle corresponding to the ($1 \sigma$)
angular resolution of the instrument
$\sigma_\theta (E)$ at energy $E$; and $t_0$ is the exposure time.
For characteristic  values
$\Delta E =(\delta E/E) E \simeq 0.25 E$, $\kappa_{\rm e.-m.}=0.5$,
$\kappa_\theta=0.68$, we obtain
$$\left(\frac{{\rm d}J_\gamma}{{\rm d} E}\right)_{\rm min}
\simeq 1.5 \cdot 10^{-9} E^{-2.1}
\left(\frac{\sigma_\theta(E)}{0.1^{\circ}}\right)
\left(\frac{S_{\rm eff}(E)}{10^{4} \, \rm m^2}\right)^{-1/2}
\left(\frac{t_0}{100 \, \rm h}\right)^{-1/2} \, \, \rm ph/cm^2  \, s \,  GeV.
$$

Our preliminary calculations show that with a stereoscopic system
consisting of several sub-100 GeV threshold IACTs, we may expect 
$\sigma_\theta \sim 0.2^{\circ}$, and 
$S_{\rm eff} \sim  4 \times 10^{4} \, \rm m^2$ at 20~GeV which results in the
integral flux sensitivity 
$J_\gamma(\geq 20 \, \rm GeV) \sim 5 \times 10^{-11} \, \rm ph/cm^2 s$.

Thus, a few sub-100 GeV imaging telescopes being combined in a
stereoscopic system could effectively intervene the domain of
satellite-based gamma ray instruments like GLAST (Bloom 1996).
But, of course,  the scientific goals of these instruments are quite different.
If the GLAST with its large (almost $2 \pi$) FoV  could provide very effective
sky surveys,  sub-100 GeV ground based instruments have an obvious advantage
for the search and study of variable sources.
On  the other side, sub-100 GeV IACT systems could operate also in the 
TeV domain by observing
the $\gamma$-ray sources at large zenith angles. If high efficiency of this
technique,  that already has been
demonstrated by the CANGAROO group at multi-TeV energies,
could be extrapolated to the 0.1-1 TeV region, the use of an
array consisting of few  10~GeV class IACT systems
(with 3 or 4 telescopes in each) could cover very broad energy
range extending  from 10 GeV to 1 TeV. Note however that more detailed  
Monte Carlo studies are needed for definite conclusions about the 
performance of sub-100~GeV stereoscopic IACT systems.  

A possible arrangement and operation modes 
of an array of 10~GeV class telescopes are described in Fig.~10.

\section{Summary}

The stereoscopic approach in the imaging atmospheric Cherenkov technique
provides superior gamma/hadron separation power ($\kappa_{\rm h} \leq 10^{-2}$),
excellent angular resolution ($\sigma_\theta \leq 0.1^{\circ}$),
good energy resolution ($\delta E/E \sim 20 \%$), and impressive flux sensitivity
($\sim 10^{-13} \, \rm erg/cm^2 s$) that makes the multi telescope arrays
as very effective tools for the study of the sky in $\gamma$-rays in a broad
energy region from 10 GeV to $\geq 10 \, \rm TeV$.
The recent results obtained by the HEGRA system of imaging telescopes, and by  
the prototype of the Telescope Array (Teshima 1997) 
generally confirm the early predictions
concerning the performance of the technique  in the TeV regime. The 
forthcoming HESS and VERITAS arrays of 10~m class imaging telescopes, with energy threshold
of about 100~GeV or less, will improve the flux sensitivities
of current instruments by an order of magnitude, and will enable 
an effective study of distant extragalactic sources with redshifts up to $z \sim 1$.
With development of new technologies of construction of 20~m class
optical reflectors and novel high quantum efficiency ($\geq 50 \%$)
optical detectors, it will be possible to extend the energy domain of the
stereoscopic IACT arrays down to 10~GeV. However, it  should be emphasized that 
if the performance of the 100~GeV threshold IACT arrays, and
their practical implementation can be predicted with high degree of confidence,
the realization of 10~GeV threshold IACT arrays still remains as an
exciting challenge.

\vspace{2mm}

We thank W.~Hofmann, A.V~Plyasheshnikov, and H.J.~V\"olk
for many fruitful discussions.

\begin{refs}

{\protect \small

\item Aharonian, F.A., Chilingarian, A.A., Mirzoyan, R.G., Konopelko A.K., 
Plyasheshnikov, A.V. 1993, Experimental Astronomy, 2, 331

\item Aharonian, F.A. (HEGRA collaboration) 1993, in Proc. ``Towards a Major
Atmospheric Cherenkov Detector-IV'' (Calgary), ed. R.C. Lamb, p.81 

\item Aharonian, F.A., Heusler, A., Hofmann, W., 
Wiedner, C.A., Konopelko A.K., Plyasheshnikov, A.V., Fomin, V.   1995, 
J.Phys.. G: Nucl. Part. Phys., 21, 985
 
\item Aharonian, F.A. (HEGRA collaboration) 1997, in Proc. 4th Compton Symposium, 
eds. C.D. Dermer et al., AIP Conf. 410, p.1631
 
\item Aharonian, F.A., Hofmann, W., Konopelko, A.K., and 
V\"olk, H.J.  1997, Astroparticle Physics, 6, I: p. 343; II: p.369

\item Aharonian, F.A., and Akerlof, C.A. 1997, Annual Rev. Nucl. Part. Sci., 47, 273   

\item Bloom, E.D. 1996, Space Sci. Rev., 75, 109

\item Cawley, M.F., and Weekes, T.C. 1996, Experimental Astronomy, 6, 7

\item Chadwick, P.M. et al. 1996, Space Sci. Rev., 75, 153

\item Daum, A. 1997, these proceedings

\item Daum, A. et al. 1997, Astroparticle Physics, in press

\item Fegan, D.J. 1997, J.Phys.. G: Nucl. Part. Phys., 23, 1013 

\item Gaidos, J.A. et al. 1996, Nature, 383, 319  

\item Goret, P. et al. 1997, in Proc. 25th ICRC (Durban), 3, p.173 

\item Hermann, G. 1997, these proceedings

\item Hillas, A.M. 1985, in Proc 19th ICRC (La Jolla), 3, p.445 

\item Hillas, A.M. 1996, Space Sci. Rev., 75, 17

\item Hofmann, W. 1997, these proceedings

\item Kifune, T. 1997, these proceedings

\item Konopelko, A.K. 1995, in Proc. ``Towards a Major
Atmospheric Cherenkov Detector-IV'' (Padova), ed. M.Cresti, p.373

\item Konopelko, A.K. 1997, these proceedings 

\item Krennrich, F., and Lamb, R.C. 1995, Experimental Astronomy, 6, 285

\item Lamb, R.C. et al. 1995a, in Proc. ``Towards a Major
Atmospheric Cherenkov Detector-IV'' (Padova), ed. M.Cresti, p.386 

\item Lamb, R.C. et al.  1995b, in Proc. ``Particle and
Nuclear Astrophysics and Cosmology in the Next Millennium'',
ed. E.W. Kolb and R.D. Peccei (World Scientific, Singapore), p.295 

\item Lorenz, E. 1997, these proceedings

\item Mirzoyan, R. 1997, these proceedings

\item M\"uller, D. et al. 1997, in Proc. 25th ICRC (Durban), 4, p.237

\item Nishimura, J. et al. 1980, ApJ, 238, 394

\item Ong, R.A. et al. 1997, these proceedings

\item Ormes, J.F. et al. 1997, in Proc. 25th ICRC (Durban), 5, p.73

\item Par\`{e} E. et al. 1997, these proceedings

\item Plyasheshnikov, A.V. and Konopelko, A.K. 1990,
 in Proc 21st ICRC (Adelaide), 4, p.250

\item Stepanian, A.A. et al. 1983, Izv. Krym. Astrophyz. Obs., 66, 234

\item Stepanian, A.A. 1995, Nuclear Physics B, 39A, 207

\item Tanimori, T. et al. 1997, ApJ, in press 

\item Teshima, M. 1997, private communication

\item Torii, S. et al. 1997,  in Proc. 25th ICRC (Durban), 4, p.241

\item Weekes, T.C., and Turver, K.E. 1977, in Proc. 12th ESLAB Symposium 
(Frascati), p.279

\item Weekes, T.C., Aharonian, F.A., Fegan, D.J., and Kifune, T. 1997a, 
in Proc. 4th Compton Symposium, eds. C.D. Dermer et al., AIP Conf. 410, p. 361

\item Weekes T.C. et al. 1997b,  in Proc. 25th ICRC (Durban), 5, p.173  

\item Zyskin, Yu. L., Stepanian, A.A., and Kornienko A.P. 1994,
J.Phys.. G: Nucl. Part. Phys., 20, 1851

}
\end{refs}

\end{document}